\documentclass[12pt]{amsart}
\hsize \textwidth      
\advance \hsize by -\marginparwidth
\evensidemargin \oddsidemargin  
\usepackage{amssymb}
\advance\hoffset by 5mm 

\newtheorem{thm}{Theorem}
\newtheorem{prop}{Proposition}
\newtheorem{lemma}{Lemma}
\newtheorem{definition}{Definition}
\newtheorem{corollary}{Corollary}

\newcommand{\pdr}[2]{\frac{\partial{#1}}{\partial{#2}}}
\newcommand{\intl}{\int\limits}
\begin{document}
\title{Quenching of  flames by fluid advection}
\author{Peter Constantin, Alexander Kiselev \and Leonid Ryzhik}
\address{ Department of Mathematics \\  
University of Chicago \\
Chicago IL 60637}

\thanks{email: const@cs.uchicago.edu, kiselev@math.uchicago.edu,
ryzhik@math.uchicago.edu} 

\begin{abstract}
  We consider a simple scalar reaction-advection-diffusion equation
  with ignition-type nonlinearity and discuss the following question:
  What kinds of velocity profiles are capable of quenching any given
  flame, provided the velocity's amplitude is adequately large? Even
  for shear flows, the answer turns out to be surprisingly subtle.
  
  If the velocity profile changes in space so that it is nowhere
  identically constant, (or if it is identically constant only in a
  region of small measure) then the flow can quench any initial data.
  But if the velocity profile is identically constant in a sizable
  region, then the ensuing flow is incapable of quenching large enough
  flames, no matter how much larger is the amplitude of this velocity.
  The constancy region must be wider across than a couple of laminar
  propagating front-widths.
  
  The proof uses a linear PDE associated to the nonlinear problem and
  quenching follows when the PDE is hypoelliptic. The techniques used
  allow the derivation of new, nearly optimal bounds on the speed of
  traveling wave solutions.
\end{abstract}
\maketitle
\section{Introduction}
We consider a mixture of reactants interacting in a region that may have
a rather complicated spatial structure but is thin across. 
A mathematical model
that describes a chemical reaction in a fluid is a system of two
equations for concentration $n$ and temperature $T$ of the form
\begin{eqnarray}\label{gen-sys}
 && T_t+u\cdot\nabla T=\kappa\Delta T+\frac{v_0^2}{\kappa}g(T)n\\
&&  n_t+u\cdot\nabla n=\frac{\kappa}{\hbox{Le}}\Delta n-
\frac{v_0^2}{\kappa}g(T)n.\nonumber
\end{eqnarray}
The equations (\ref{gen-sys}) are coupled to the reactive Euler equations
for the advection velocity $u(x,y,t)$.  Two assumptions are usually
made to simplify the problem: the first is a constant density approximation
\cite{CW} that allows to decouple the Euler equations from the system
(\ref{gen-sys}) and to consider  $u(x,y,t)$ as a prescribed
quantity that does not depend on $T$ and $n$.  The second assumption  is
that $\hbox{Le}=1$  (equal  thermal and
material diffusivities). These two assumptions 
reduce the above system to a single scalar equation for the
temperature $T$. We assume in addition that the advecting flow is
unidirectional. Then the system (\ref{gen-sys}) becomes
\begin{eqnarray}
  \label{eq:1.1}
  &&T_t+Au(y)T_x=\kappa\Delta T+\frac{v_0^2}{\kappa}f(T)\\
&& T(0,x,y)=T_0(x,y)\nonumber
\end{eqnarray}
with $f(T)=g(T)(1-T)$. We are interested in strong advection, and accordingly 
have written the velocity as a product of the amplitude $A$ and the profile
$u(y)$. In this paper we consider nonlinearity of the ignition
type
\begin{eqnarray}
\nonumber
 && \hbox{(i) $f(T)$ is Lipschitz continuous on $0\le T\le 1$},\\
&&  \hbox{(ii) }~f(1)=0,~~\exists\theta_0
\hbox{ such that $f(T)=0$ for $0\le T\le\theta_0$, $f(T)>0$ for
$T > \theta_0,$ }\label{eq:2.1.2}\\
&&\hbox{(iii) }~~f(T)\le T.\nonumber
\end{eqnarray}
The last condition in (\ref{eq:2.1.2}) is just a normalization.
We consider the reaction-diffusion equation (\ref{eq:1.1}) in a
strip $D=\left\{x\in {\mathbb R} ,~ y\in [0,H]\right\}$.
Equation (\ref{eq:1.1})
may be considered as a simple model of flame propagation in
a fluid \cite{Ber-Lar-Lions}, advected by a shear (unidirectional) flow.
The physical literature on the subject is vast, and we refer to the
recent review \cite{jxin-3} for an extensive bibliography. The main
physical effect of advection for front-like solutions is the speed-up of
the flame propagation due to the large scale distortion of the front. 
The role of the advection term in (\ref{eq:1.1}) for the front-like
initial data was also a subject of intensive mathematical scrutiny recently.
Existence of unique
front-like traveling waves 
has been established in \cite{Ber-Lar-Lions,Ber-Nir-2}, 
and their stability has been studied in \cite{Mal-Roq, Roq, jxin-2}. 
A traveling front is a solution
of (\ref{eq:1.1}) of the form 
\begin{equation}
  \label{eq:tr-w1}
T(t,x,y)=U(x-c_At,y)
\end{equation}
 such that
\begin{equation}\label{eq:tr-w2}
\lim_{s\to -\infty}U(s,y)=1,~~
\lim_{s\to +\infty}U(s,y)=0,\,\,\,
U_s(s,y) <0.
\end{equation}
The monotonicity property is not required for traveling
wave solutions, but it is always present in the situation we consider.
The speed-up of the fronts by advection mentioned above may be
quantified as the dependence of the traveling front speed $c_A$ on
the amplitude $A$.  Variational formulas for $c_A$ were derived in
\cite{Hamel, HPS}. The latter work also contains results on the
asymptotic behavior of $c_A$ when $A$ is small for some classes of
shear flows, as well as upper bounds on $c_A$ linear in $A$.  An
alternative approach to quantifying advection effects was introduced
by the present authors in \cite{CKOR,KR}. It is based on the notion of
the bulk burning rate,
\[ V(t) = \int_D T_t \, \frac{dxdy}{H} = 
\frac{v_0^2}{\kappa}\int_D f(T) \,\frac{dxdy}{H}\]
which extends the notion of  front speed.
We derived lower bounds for long time averages of $V(t)$ which behave 
like $CA$ for large $A,$ with constant $C$ depending on the geometry 
of the flow. These bounds are valid for a class of flows that  we call 
percolating. They are characterized by infinite tubes of streamlines 
connecting $\pm \infty$ and include shear flows as a particular case. 
Our bounds imply the estimate $c_A \geq CA$ for  traveling 
waves. Audoly, Berestycki and Pomeau gave a formal 
argument \cite{ABP} suggesting that in the case when the shear flow 
varies on the scale much larger than the laminar flame width, 
one should have $c_A \sim A.$ One of the by-products of this paper is a 
rigorous proof of this conjecture.    
  

Our main goal in the present paper is to 
consider advection effects for a different physically interesting situation, 
where initial data are compactly supported. In this case,
two generic scenarios are possible. If the support of the initial data
is large
enough, then two fronts form and propagate in opposite directions. 
Fluid advection speeds up the propagation, 
accelerating the burning. However,
if the support of the initial data 
is small, then the advection  exposes the initial hot region  to 
diffusion which cools it below the ignition temperature,  ultimately 
extinguishing the flame. 

We consider for simplicity  periodic boundary conditions 
\begin{equation}
  \label{eq:2.1.1}
  T(t,x,y)=T(t,x,y+H)
\end{equation}
in $y$ and decay in $x$:
\begin{eqnarray}\label{eq:decay}
T(t,x,y)\to 0~~\hbox{as $|x|\to\infty$.}
\end{eqnarray}
We take  
\begin{equation}
  \label{eq:3.1.1}
  \intl_{0}^Hu(y)dy=0.
\end{equation}
A constant non-zero mean can be easily taken into account by translation.
We consider the  case when the width of the  domain is larger than the
laminar front width length scale: $H\gg l =\frac{\kappa}{v_0}.$ 
We will always assume that initial data $T_0(x,y)$ is such that $0\le
T_0(x,y)\le 1$. Then we have $0\le T\le 1$ for all $t>0$ and $(x,y)\in
D$. Moreover, we assume that for some $L$ and $\eta>0$ we have
\begin{eqnarray}\label{eq:2.1.4}
&& T_0(x,y)>\theta_0+\eta~~\hbox{for $|x|\le L/2$},\\
&& T_0(x,y)=0~~\hbox{for $|x|\ge L$}. \nonumber 
\end{eqnarray}

The main purpose of this paper is to study the possibility of
quenching of flames by strong fluid advection in a model
(\ref{eq:1.1}). The phenomena associated with flame quenching are of
great interest for physical, astrophysical and engineering
applications. The problem of extinction and flame propagation in the
mathematical model (\ref{eq:1.1}) under conditions (\ref{eq:2.1.2}),
(\ref{eq:decay}), (\ref{eq:2.1.4}), was first studied by Kanel
\cite{Kanel'} in one dimension and with no advection. He showed that,
in the absence of fluid motion, there exist two length scales
$L_0<L_1$ such that the flame becomes extinct for $L<L_0$, and
propagates for $L>L_1$.  More precisely, he has shown that there exist
$L_0$ and $L_1$ such that
\begin{eqnarray}
  \label{eq:2.2}
&&  T(t,x,y)\to 0~\hbox{as $t\to\infty$ uniformly in $D$ if $L<L_0$}\\
&&T(t,x,y)\to 1~\hbox{as $t\to\infty$ for all $(x,y)\in D$ if $L>L_1$}.
\nonumber
\end{eqnarray}
In the absence of advection, the flame extinction is achieved by
diffusion alone, given that the support of initial data is small compared to the
scale of the laminar flame width $l = \kappa/v_0.$ However, in many
applications the quenching is the result of strong wind, intense fluid
motion and operates on larger scales. There are few results available for such situations
in the framework of the reaction-diffusion model.  Kanel's result was
extended to non-zero advection by shear flows by Roquejoffre
\cite{Roq-2} who has shown that (\ref{eq:2.2}) holds also for $u\ne 0$
with $L_0$ and $L_1$ depending, in particular, on $A$ and $u(y)$. The
second (propagation) part in (\ref{eq:2.2}) was also proved in
\cite{jxin-2} for general periodic flows. However the interesting
question about more explicit quantitative dependence of $L_0,$ $L_1$
on $A$ and $u(y)$ remained open. Is it possible to
quench the initial data that previously lead to an expanding solution
by increasing $A,$ but not changing the profile?
How does this possibility depend on geometry of the profile $u(y)?$ Anyone
who has tried to light a match in the wind has some intuition about
this phenomenon. Yet, the mathematical answer turns out to be
surprisingly subtle.  In this paper, we also limit ourselves to shear
flows.  We are interested in the understanding the behavior of $L_0$
and $L_1$ for large $A.$ The answer  depends
strongly on the geometry of the flow. In  some
cases the maximal extinction size grows $L_0\sim A$, and in others even the
propagation size $L_1$ remains finite as $A$ goes to infinity.  In the
first case, we will say that $u(y)$ is quenching. 
\begin{definition}
  We say that the profile $u(y)$ is quenching if for any $L$ and any
  initial data $T_0(x,y)$ supported inside the interval
  $[-L,L]\times[0,H],$ there exists $A_0$ such that solution of
  (\ref{eq:1.1}) becomes extinct:
\[
T(t,x,y)\to 0~\hbox{as $t\to\infty$ uniformly in $D$}
\]
for all $A\ge A_0$.  We call the profile $u(y)$ strongly quenching if the
critical amplitude of advection $A_0$
satisfies $A_0 \leq C L$ for some constant $C(u,\kappa,v_0,H)$ (which
has the dimension of inverse time).
\end{definition}
The key feature that distinguishes quenching from non-quenching
velocities is the absence or presence of large enough flat parts in the
profile $u(y)$.
\begin{definition} 
  We say that the profile $u(y)\in C^\infty[0,H]$ satisfies the
  H-condition if 
 \begin{equation}
    \label{eq:hormander}
\hbox{
there is no point $y\in[0,H]$, where all
  derivatives of $u(y)$ vanish.}
\end{equation}
\end{definition}
The H-condition guarantees that the operator 
\begin{equation}\label{eq:hypo}
u(y)\pdr{}{x}-\frac{\partial^2}{\partial y^2}
\end{equation}
is hypoelliptic \cite{Hormander}. The study of existence of smooth fundamental
solutions for such operators was initiated by Kolmogorov
\cite{Kolmogorov}.  Kolmogorov's work with $u(y)=y$ served in part
as a motivation for the fundamental result on characterization of 
hypoelliptic operators of H\"ormander \cite{Hormander}.
The hypoellipticity of the operator
(\ref{eq:hypo}) plays a key role in our considerations.  Our first
result is that the H-condition implies strong quenching.
\begin{thm}\label{thm1}
  Let $f(T)$ be an ignition type nonlinearity, and let $u\in
  C^{\infty}[0,H]$ satisfy the H-condition. Then $u(y)$ is strongly
  quenching.  That means that there exists a constant $C(u, \kappa,
  v_0, H) >0$ that may depend on $H$, $\kappa$, $v_0$ and $u(y)$ but
  is independent of $A$ such that
  \begin{equation}
    \label{eq:thm1}
\hbox{$T(t,x,y)\to 0$ as
  $t\to\infty$ uniformly in $D$.}
\end{equation}
This flame extinction occurs whenever the initial temperature
$T_0(x,y)$ is supported in an interval $[-L,L]\times[0,H],$ with
$L<C(u,\kappa,v_0,H) A$ .
\end{thm} 

The next result shows that a plateau 
on the order of the laminar front width 
\begin{equation}
l=\kappa/v_0\label{ell}
\end{equation}
in the profile $u(y)$ prohibits
quenching. (And therefore the conditions in Theorem~\ref{thm1} are natural.)

\begin{thm}\label{thm2} There exist universal constants $C_0, C_1>0$,
  such that, if  $u(y)=\bar u=\hbox{const}$ for $y\in[a-h,a+h]$
  for some $a\in[0,H]$ and $h\ge h_0=C_0 l$, 
 then  
  \begin{eqnarray}
    \label{eq:1.2}
    T(t,x,y)\to 1~~ \hbox{as $t\to \infty$}
  \end{eqnarray}
uniformly on compact sets, for all $A\in {\mathbb R} $.  This flame
propagation occurs whenever
the initial temperature $T_0(x,y)$ satisfying  (\ref{eq:2.1.4})
is supported in an interval $[-L,L]\times[0,H]$ with  $L\ge C_1 h_0$. 
\end{thm}

An interesting by-product of the proof of Theorem~\ref{thm2}  
is an estimate for the speed of traveling front solutions of 
(\ref{eq:1.1}) when the shear flow varies slowly on the scale
of the laminar flame width $l.$
Let us define
\[
|u_+|_{h_0}=\max_{0\le y_0\le H}\left\{\min_{y\in[y_0-h_0,y_0+h_0]}u_+\right\}
\]
with $h_0=Cl$ given by Theorem \ref{thm2}, and $u_+ = {\rm max}(u(y),0).$
\begin{thm}\label{thm3}
The speed of the traveling front $c$ satisfies the upper and lower bounds
\[
A|u_+|_{h_0}\le c\le A\|u_+\|_{\infty}+v_0.
\] 
\end{thm}
The upper bound of Theorem~\ref{thm3} is contained in \cite{CKOR} (it
is shown there for KPP type reaction, but this immediately implies the
corresponding bound for ignition nonlinearity by a simple application
of maximum principle). The left hand side is close to
$\|u_+\|_{\infty}$ if $u(y)$ is slowly varying on the scale $h_0$.
This agrees with the formal prediction of Audoly, Berestycki and
Pomeau \cite{ABP}, and also (up to the addition of $v_0$) with the
results of Majda and Souganidis \cite{MS} in the homogenization regime
$\kappa\to 0$.

Unlike hypoellipticity,  the quenching property is stable to small $L^{\infty}$ perturbations: a small enough  plateau (on the scale of the laminar front
width $l$) does not stop quenching.
\begin{thm}\label{thm4}
  For every $\theta_0>0$ in (\ref{eq:2.1.2}) there exists a constant $B>0$ such
  that, if a profile $u(y)$ satisfies the H condition outside
  an interval $y\in[a-h,a+h]$ with  $h\le h_1=Bl$,  then it is 
strongly quenching.
\end{thm}
Moreover, the strongly quenching profiles are generic in the following sense:
\begin{thm}\label{thm5}
The set of all strongly quenching shear flows 
$u(y)$ contains a dense open set in  $C[0,H]$
(here $C[0,H]$ is the space of continuous functions on $[0,H]$).
\end{thm} 

In Section~\ref{sys}  we show that
all results on quenching, namely, Theorems~\ref{thm1}, \ref{thm4}, 
\ref{thm5} extend to the case of full system (\ref{gen-sys}) with 
$\hbox{Le} \ne 1.$

Finally, in the last section we prove that initial data of sufficiently 
small size (of the same order as in the case $u(y)=0$) will be 
quenched by any shear flow $Au(y).$


{\bf Acknowledgment.} We thank N. Nadirashvili for fruitful
discussions, and in particular for suggesting that we look for 
radially symmetric sub-solutions. We also thank G. Papanicolaou for 
stimulating discussions.  This work
partially supported by ASCI Flash Center at the University of
Chicago. PC was supported by NSF grant DMS-9802611, AK was supported by
NSF grant DMS-9801530, and LR was supported by NSF grant DMS-9971742.
 
\section{Quenching by a shear flow}

We prove Theorem \ref{thm1} in this section. 
\begin{proof}[Proof of Theorem \ref{thm1}]
It suffices to show that
there exists some time $t_0$ such that 
\begin{equation}\label{eq:smallT}
\hbox{$T(t_0,x,y)\le\theta_0$ for all
$(x,y)\in D$.}
\end{equation}
Then it follows from the maximum principle that
$T(t,x,y)\le\theta_0$ for all $t\ge t_0$, and hence $T$ satisfies the
linear advection-diffusion equation 
\[
T_t+Au(y)T_x=\kappa\Delta T.
\]
for $t\ge t_0$, which implies (\ref{eq:thm1}). We actually show that
(\ref{eq:smallT}) holds at $t_0=\kappa/v_0^2$ for sufficiently large $A$.
Recall that $f(T)\le
T$, and hence $T(t,x,y)$ can be bounded from above using the maximum
principle as follows:
\begin{equation}
  \label{eq:2.3}
  T(t,x,y)\le \Phi(t,x,y)e^{{v_0^2}t/\kappa}.
\end{equation}
Here the function $\Phi(t,x,y)$ satisfies the linear problem
\begin{eqnarray}
\label{eq:2.4}
  &&\Phi_t+Au(y)\Phi_x=\kappa\Delta\Phi\\
&&\Phi(0,x,y)=T_0(x,y)\nonumber\\
&&\Phi(t,x,y)=\Phi(t,x,y+H).\nonumber
\end{eqnarray}
Furthermore, we have 
\[
\Phi(t,x,y)=\int_{-\infty}^\infty dzG(t,x-z)\Psi(t,z,y)
\]
with the function $\Psi(t,x,y)$ satisfying the degenerate parabolic equation
\begin{eqnarray}
  \label{eq:2.5}
  &&\Psi_t+Au(y)\Psi_x=\kappa\Psi_{yy}\\
&&\Psi(0,x,y)=T_0(x,y)\nonumber\\
&&\Psi(t,x,y)=\Psi(t,x,y+H)\nonumber
\end{eqnarray}
and
\[
G(t,x)=\frac{1}{\sqrt{4\pi\kappa t}}\exp\left(-\frac{x^2}{4\kappa t}\right).
\]
We note that if $u(y)$ satisfies the H-condition (\ref{eq:hormander})
then the diffusion process defined by (\ref{eq:2.5}) has a unique
smooth transition probability density. Indeed, the Lie algebra
generated by the operators $\partial_y$ and
$\partial_t+u(y)\partial_x$ consists of vector fields of the form
\[
\pdr{}{y},\pdr{}{t}+u(y)\pdr{}{x},
u'(y)\pdr{}{x},u''(y)\pdr{}{x},\dots,u^{(n)}(y)\pdr{}{x},\dots
\]
which span $ {\mathbb R} ^2$ if $u(y)$ satisfies (\ref{eq:hormander}).
Then the theory of H\"ormander \cite{Hormander}, and the results of
Ichihara and Kunita \cite{kunita} imply that there exists a smooth
transition probability density $p_A(t,x,y,y')$ such that
\[
\Psi(t,x,y)=\int\limits_{{\mathbf R}}  dx'\int\limits_{0}^Hdy'
p_A\left(t,x-x',y,y'\right)T_0(x',y').
\]
In particular, the function $p_A(t)$ is uniformly bounded from above for
any $t>0$ \cite{kunita}.  Then we have
\begin{eqnarray*}
  T(t,x,y)\le e^{v_0^2t/\kappa}\|\Phi(t)\|_{L^\infty_{x,y}}\le 
e^{v_0^2t/\kappa}
\|\Psi(t)\|_{L^\infty_{x,y}}\le e^{v_0^2t/\kappa}
{\|p_A(t)\|_{L^\infty_{x,y}}}
\|T_0\|_{L^1_{x,y}}.
\end{eqnarray*}
It is straightforward to observe that 
\[
p_A(t,x,y,y')=\frac{v_0}{A}p_0(t,\frac{A}{v_0}x,y,y')
\]
with $p_0$ being the transition probability density for (\ref{eq:2.5}) with
$A=v_0$. That is, $p_0$ satisfies
\begin{eqnarray*}
 && \pdr{p_0}{t}+v_0u(y)\pdr{p_0}{x}=\kappa
\frac{\partial^2 p_0}{\partial y^2}\\
 &&p_0(0,x,y,y')=\delta(x)\delta(y-y')\\
&&p_0(t,x,y)=p_0(t,x,y+H).
\end{eqnarray*}
Therefore we obtain
\[
   T(t,x,y)\le e^{v_0^2 t/\kappa}\frac{v_0}{A}\|p_0(t)\|_{L_{x,y}^\infty}
\|T_0\|_{L_{x,y}^1}\le 4e^{v_0^2 t/\kappa} 
\frac{v_0}{A}\|p_0(t)\|_{L_{x,y}^\infty}LH
\]
and in particular at time $t_0=\frac{\kappa}{v_0^2}$ we have
\begin{eqnarray}\label{eq:below}
   T(t_0,x,y)\le C\frac{v_0}{A}\|p_0(t_0)\|_{L_{x,y}^\infty}LH
\le \theta_0
\end{eqnarray}
as long as
\begin{eqnarray*}
  \frac{A}{v_0}\ge C\frac{\|p_0(t_0)\|_{L_{x,y}^\infty}LH}{\theta_0}. 
\end{eqnarray*}
Theorem \ref{thm1} follows from (\ref{eq:below}) as explained in the
beginning of this Section.  
\end{proof}

We prove now Theorem \ref{thm4} that shows that a sufficiently small
plateau in the profile $u(y)$ is not an obstruction to quenching.

\begin{proof}[Proof of Theorem \ref{thm4}]  Let
us define the set
\[
D_r= {\mathbb R} \times [a-r,a+r].
\]
As before, it suffices to show that solution of (\ref{eq:2.4}) satisfies
\begin{equation}
  \label{eq:4.0}
  \Phi(t_0,x,y)\le \frac{\theta_0}{e},~~t_0=\frac{\kappa}{v_0^2}
\end{equation}
and that is what we will do.  First, we split the initial data for
(\ref{eq:2.5}) into two parts: one supported on a strip $D_{h_1}$,
containing the flat part of $u(y),$ and another supported outside it.
We will choose $h_1=C_1 l>h$ such that any solution of (\ref{eq:2.5})
that is independent of $x$ and with initial data supported inside
$D_{h_1}$ will be small at time $t_0=\kappa/v_0^2$. The second part is
supported away from the strip $D_h$, where $u(y)$ is flat. Therefore
for a sufficiently small time it behaves like a solution of
(\ref{eq:2.5}) with advection satisfying the H-condition.

We choose $h_1$ as follows. Let $\phi(t,y)$ be a periodic solution of
\[
\phi_t=\phi_{yy},~~\phi(0,y)=\phi_0(y),~~~0\le\phi_0(y)\le 1
\]
given by
\[
\phi(t,y)=\sum_{j\in Z}\phi_j(0)e^{2ij\pi y/H-4\kappa t\pi^2j^2/H^2}.
\]
Then we have
\[
\left|\phi(t,y)-\frac{\|\phi(0,y)\|_{L_y^1}}{H}\right|\le 
2\frac{\|\phi(0,y)\|_{L_y^1}}{H}\sum_{j\ge 1}
e^{-4\kappa t\pi^2j^2/H^2}.
\]
A simple estimate shows that 
\[ \sum_{j\ge 1}
e^{-\frac{4\kappa^2\pi^2j^2}{v_0^2 H^2}} \leq C \frac{Hv_0}{\kappa},   \]
and hence
\begin{equation}\label{eq:simpleest}
\left|\phi(t_0,y)-\frac{\|\phi(0,y)\|_{L_y^1}}{H}\right|\le C 
\frac{\|\phi(0,y)\|_{L_y^1}}{l},~~t_0=\frac{\kappa}{v_0^2}.
\end{equation}
Therefore we have
\[  \phi(t_0,x,y)\le\frac{\theta_0}{10}
\]
as long as 
\begin{equation}
  \label{eq:4.2}
  \|\phi(0,y)\|_{L_y^1}\le C_1 l,
\end{equation}
where $C_1= \theta_0 Hl / (10(l+CH)).$ 
Let us choose $h_1$ so that (\ref{eq:4.2}) is automatically verified 
for initial data supported on $D_{2h_1}:$ 
\[
  h_1 \le C_1 l/4
\]
with $C_1$ as in (\ref{eq:4.2}). Let us assume that the width of the interval 
$[a-h,a+h]$, on which $u(y)$ is constant satisfies
\[
  h\le \frac{h_1}{4}.
\]
This condition determines the constant $B$ in the statement of
Theorem~\ref{thm4}.  We may now split the initial data for
(\ref{eq:2.5}) as follows:
\[
T_0(x,y)\le \chi_0(y)+\psi_0(x,y).
\]
Here the smooth function $\chi_0(y)$ is supported in the interval
$[a-2h_1,a+2h_1]$ while the function $\psi_0(x,y)$ is supported
outside the set $[a-h_1,a+h_1]$. Both of these functions satisfy in
addition $0\le\chi_0(y),\psi_0(x,y)\le 1$. Then the function
$\Phi(t,x,y)$ satisfies the inequality
\[
\Phi(t,x,y)\le \chi(t,y)+\psi(t,x,y)
\]
with the functions $\chi$ and $\psi$ satisfying (\ref{eq:2.5}) with
the initial data $\chi_0$ and $\psi_0$, respectively. It follows from
our choice of $h_1$ that
\[
  \chi(t_0,y)\le\frac{\theta_0}{10},
\]
so it remains only to estimate $\psi(t_0,x,y)$. We will do it
separately for $(x,y)$ inside and outside of the strip $D_{h_1/2}$.
For the points $(x,y)\in D_{h_1/2}$ we have:
\begin{eqnarray*}
  &&\psi(t,x,y)=\intl_{{\mathbf R}}dx'
\intl_{0}^Hdy' p_A(t,x-x',y,y')\psi_0(x',y')\le
\intl_{ {\mathbf R} }dx'\intl_{|y'-a|\ge h_1}dy'p_A(t,x-x',y,y')\\
&&\le {\mathcal P}\left\{|W(t)|\ge \frac{h_1}{2}\right\}\le\frac{\theta_0}{10}
\end{eqnarray*}
for sufficiently small $t$. Here $W(t)$ is the one-dimensional
Brownian motion with diffusivity $\kappa$, and ${\mathcal P}$ denotes
probability with respect to it. Thus (\ref{eq:4.0}) holds inside $D_{h_1/2}$.

In order to estimate the function
$\psi(t,x,y)$ outside $D_{h_1/2}$ we introduce a profile $\tilde u(y)$
that coincides with $u(y)$ outside of the interval
$[a-(1+\delta)h,a+(1+\delta)h]$, $\delta\ll 1$, and satisfies the
H-condition on the whole interval $[0,H]$. We define the
process $(X(t),Y(t))$ by
\begin{equation}\label{eq:4.7}
dX(t)=u(Y(t))dt,~~dY(t)=\sqrt{2\kappa} dW(t),~~X(0)=x,~~Y(0)=y.
\end{equation}
Consider the stopping time $\tau$ which is the first time when
$Y(t)$ enters the interval $[a-(1+\delta)h,a+(1+\delta)h]$.
Then we have
\begin{eqnarray*}
&&\psi(t,x,y)\le P_{x,y}\left\{(X(t),Y(t))\in\hbox{supp}~\psi_0\right\}
=P_{x,y}\left\{(X(t),Y(t))\in\hbox{supp}~\psi_0|\tau >t\right\}
P_y(\tau >t)\\
&&+P_{x,y}\left\{(X(t),Y(t))\in\hbox{supp}~\psi_0|\tau <t\right\}
P_y(\tau<t)\\
&&\le 
P_{x,y}\left\{(\tilde X(t),\tilde Y(t))\in\hbox{supp}~\psi_0|\tau >t\right\}
P_y(\tau >t)+P_y(\tau<t).
\end{eqnarray*}
Here $P_{x,y}$ denotes probability with respect to the process
$(X(t),Y(t))$ starting at $(x,y)$, while $P_y$ denotes probability
with respect to $Y(t)$ starting at $y$ (recall that $Y(t)$ is
independent of $x$).  The
process $X(t)$ for $t<\tau$ is identical to the process $(\tilde
X(t),\tilde Y(t))$ defined by (\ref{eq:4.7}) with $u(Y)$ replaced by
$\tilde u(Y)$.  Therefore we have
\begin{eqnarray*}
 &&\psi(t,x,y)\le P_{x,y}\left\{(\tilde X(t),\tilde Y(t))
\in\hbox{supp}~\psi_0|\tau >t\right\}
P_y(\tau >t)+P_y(\tau<t)\\
&&\le  P_{x,y}\left\{(\tilde X(t),\tilde Y(t))
\in\hbox{supp}~\psi_0\right\}+P_y(\tau<t).
\end{eqnarray*}
Recall that $(x,y)\notin D_{h_1/2}$ and $h\le h_1/4$. Therefore the
point $y$ is a fixed distance away from the interval
$[a-(1+\delta)h,a+(1+\delta)h]$. Hence we may choose $t_1<t_0$
sufficiently small so that
\[
P_y(\tau<t_1)\le\frac{\theta_0}{10}.
\]
Furthermore, the function $\tilde\psi(t,x,y)=
P_{x,y}\left\{(\tilde X(t),\tilde Y(t))\in\hbox{supp}~\psi_0\right\}$ satisfies
(\ref{eq:2.5}) with the initial data
\[
\tilde\phi(0,x,y)=\left\{\begin{array}{ll} 1, & (x,y)\in {\rm supp}~\psi_0 \\
0, & (x,y)\notin{\rm supp}~\psi_0 \end{array}\right.
\]
However, $\tilde u(y)$ is quenching and thus we may choose $A$ so large that
\[
\tilde\psi(t_1,x,y)\le\frac{\theta_0}{10}.
\]
Therefore we have at $t=t_1$:
\[
\psi(t_1,x,y)\le\frac{\theta_0}{5}
\]
and hence the same upper bound holds at $t=t_0>t_1$.  Therefore
(\ref{eq:4.0}) holds also outside $D_{h_1/2}$, and Theorem \ref{thm4}
follows. The fact that $u(y)$ is strongly quenching follows from 
this property of $\tilde u(y).$ 
\end{proof}

Theorem~\ref{thm5} is a simple corollary of Theorem~\ref{thm1}. \\
\begin{proof}[Proof of Theorem~\ref{thm5}]
  The set of all profiles $u(y)$ satisfying H-condition is dense in
  $C[0,H],$ so by Theorem~\ref{thm1} the set of strongly quenching
  profiles is dense. To complete the proof, we will show that if
  $\tilde u(y)$ satisfies H-condition, then there exists
  $\delta(\tilde u)$ such that if
\[ \|u(y)-\tilde u(y)\|_{C[0,H]} <\delta(\tilde u) \]
then $u(y)$ is strongly quenching.  From the proof of
Theorem~\ref{thm1}, we know that there exists a constant $C(\tilde u)$
such that the solution $\tilde \Psi(x,y,t)$ of the equation
(\ref{eq:2.5}) with advection $A\tilde u(y)$ satisfies
\[ \tilde \Psi(x,y,t_0) \leq \frac{\theta_0}{e} \]
if the initial data $\tilde \Psi(x,y,0)$ is supported on the interval
$[-L,L]$ with $L < C(\tilde u)A$ (recall that such inequality
implies quenching of $T$ with the same initial data).  Let $\Psi(x,y,t)$ be
a solution of (\ref{eq:2.5}) with advection $Au(y),$ and initial data
supported in $[-L',L'].$ Then by the Feynman-Kac formula (see, e.g.
\cite{Freidlin})
\begin{eqnarray*}
\Psi(x,y,t)\leq P_{x,y}(x+A\int\limits_0^t u(y+W(s))\,ds \in [-L',L'])  \leq \\
P_{x,y}(x+A\int\limits_0^t \tilde u(y+W(s))\,ds 
\in [-L'-A\delta t,L'+A\delta t]) \le
\tilde \Psi(x,y,t),  
\end{eqnarray*}
where $\tilde \Psi(x,y,t)$ is the solution of (\ref{eq:2.5}) with
advection $Au(y)$ and initial data equal to the characteristic
function of the interval $[-L'-A\delta t_0, L'+A\delta t_0]$ for $t\le
t_0$. Now choose $\delta < C(\tilde u)/t_0$. Then
\[ \Psi(x,y,t_0) \leq \tilde \Psi(x,y,t_0) \leq \frac{\theta_0}{e} \]
provided that $L' \leq (C(\tilde u)-\delta t_0)A,$ and hence $u(y)$ is
strongly quenching.  
\end{proof}

\section{Quenching for a system}
\label{sys}

All results on quenching for equation (\ref{eq:1.1}) proved in the
previous section extend directly to the case of the system
(\ref{gen-sys}). In this respect, the situation is similar to the case
$u(y)=0$, where all results on quenching proved by Kanel \cite{Kanel'}
for a single equation extend to the system case.  We make an
assumption $g(T) \leq T,$ which is just a normalization and
corresponds to the condition $f(T) \leq T$ in the case of a single
equation.  We take compactly supported initial data for the
temperature, while for the concentration we assume $0 \leq n(x,y,0) \leq
1.$ Notice that by the maximum principle, $n(x,y,t) \leq 1$ for every $t.$
Therefore, $T$ satisfies
\[ T_t +u(y)T_x -\kappa \Delta T \leq T, \]
and from this point the analysis proceeds in the same way as for 
scalar equation (starting from (\ref{eq:2.3})). We summarize the results in
\begin{thm} \label{thm6}
  Assume that $g(T)=0$ for $T \leq \theta_0,$ and $g(T) \leq T.$
  Suppose that the velocity profile $u(y) \in C_0^\infty[0,H]$
  satisfies H-condition, then it is strongly quenching. The same
  conclusion holds if the profile $u(y)$ has a plateau of the size $h
  \leq h_1=Bl$ (where $B$ is a universal constant).  Moreover, the set
  of all quenching shear flows contains a dense open set in $C[0,H]$.
\end{thm}
 
\section{Flame propagation }\label{sec:propagate}

We prove Theorem \ref{thm2} in this section.  Let $T(t,x,y)$ be solution of
(\ref{eq:1.1}) with the initial data as in (\ref{eq:decay}).  We will
use the following result of Xin \cite{jxin-2} that holds for more
general types of advection (its version for the shear flow was also
proved by Roquejoffre in \cite{Roq-2}). Consider
\begin{eqnarray}
  \label{eq:3.2.1}
  &&T_t+u(x,y)\cdot \nabla T=\kappa\Delta T+\frac{v_0^2}{\kappa}f(T)\\
&&T(0,x,y)=T_0(x,y)\in L^2(D)\nonumber
\end{eqnarray}
with $u(x,y)$ being periodic in both variables, and $0\le T_0(x,y)\le 1$.
\begin{prop}[Xin]
\label{prop1} Assume that the initial data in
  (\ref{eq:3.2.1}) are such that
\[
\lim_{|x|\to\infty}T_0(x,y)=0~~\hbox{uniformly in $D$}
\]
and
\[
T_0(x,y)>\theta_0+\eta~~\hbox{for $|x|\le L$}. 
\]
Then there exists $L_1(\eta,u)$
that depends on $\eta$ and $u(x,y)$ such that if $L\ge L_1$ then
\begin{eqnarray*}
  \lim_{t\to\infty}T(x-st,y,t)=
\left\{\begin{array}{ll}1, & \hbox{if $c_l<s<c_r$}\\
0, & \hbox{if $c_l>s$ or $s>c_r$}\end{array}\right.
\end{eqnarray*}
Here $c_l$ and $c_r$ are the speeds of left and right traveling waves,
respectively ($c_l<0$, $c_r>0$). 
\end{prop}
The right traveling wave is described by (\ref{eq:tr-w1}), (\ref{eq:tr-w2}),
and the left traveling wave satisfies (\ref{eq:tr-w1}) and
\[
\lim_{s\to -\infty}U(s,y)=0,~~
\lim_{s\to +\infty}U(s,y)=1,~~U_s(s,y)>0.
\]
Notice that Theorems \ref{thm1} and \ref{thm4} imply an estimate
$L_1\ge CA$ if $u$ has no flat parts larger than certain critical
size.  On the other hand, Theorem \ref{thm2} shows that $L_1=h_0$ is
independent of $A$ if $u$ has a sufficiently large flat part.

\begin{proof}[Proof of Theorem \ref{thm2}]
The proof of Theorem \ref{thm2} proceeds in several steps. We 
consider the initial data satisfying (\ref{eq:2.1.4}). First we
find $h_0$ such that there exists a $C^2$ function $\phi(x,y)$ such
that $0\le\phi<\theta_0+\eta$, and 
\begin{eqnarray}
  \label{eq:3.3}
  \kappa\Delta\phi+\frac{v_0^2}{\kappa}f(\phi)\ge 0
\end{eqnarray}
and $\phi$ vanishes on the boundary of the disc of radius $h_0$ centered at
the point $(0,a)$:
\begin{eqnarray*}
  \phi |_{\partial\Omega_0}=0,~~\Omega_0=B((0,a);h_0).
\end{eqnarray*}
Then in the system of coordinates that moves with the speed $\bar u$
the function $\phi(x,y)$ is a sub-solution of (\ref{eq:1.1}) in 
$\Omega_0$. Therefore, initial data that start above $\phi$ will not
decay to zero. Next we consider the special solution
$\Phi(t,x,y)$ of (\ref{eq:1.1}) with the initial data given by
\begin{equation}\label{eq:3.4.1}
\Phi_0(x,y)=\left\{\begin{array}{ll}\phi(x,y) & (x,y)\in\Omega_0 \\ 0 &
      (x,y)\notin\Omega_0\end{array}\right.
\end{equation}
We show that $\Phi(t,x,y)$ satisfies (\ref{eq:1.2}) and that 
implies that (\ref{eq:1.2}) holds for arbitrary initial data
$T_0\ge\Phi_0$, in particular, such as described in Theorem
\ref{thm2}.

{\bf Step 1. Construction of a stationary sub-solution.} 
Choose $\theta_1,\theta_2$
so that $\theta_0+\eta>\theta_2>\theta_1>\theta_0$ and define $f_1(T)$ by
\begin{eqnarray*}
  f_1(T)=\left\{\begin{array}{ll} 0, & T\le \theta_1 \\ \displaystyle 
\frac{f(\theta_2)(T-\theta_1)}{\theta_2-\theta_1}, & 
\theta_1\le T\le\theta_2 \\ 
f(T), &\theta_2\le T\le 1 \end{array}\right.
\end{eqnarray*}
The function $f(T)$ is Lipschitz continuous, and hence we may choose
$\theta_1$ and $\theta_2$ so that $f_1(T)\le f(T)$. Therefore if
$\phi$ satisfies
\begin{equation}\label{eq:3.6}
\kappa\Delta\phi+\frac{v_0^2}{\kappa}f_1(\phi)=0
\end{equation}
then $\phi$ satisfies (\ref{eq:3.3}).  We are going  to
construct an explicit radial solution $\phi(r)$ of (\ref{eq:3.6}) with the
``initial'' conditions
\begin{eqnarray*}
\phi(0)=\theta_2,~~\pdr{\phi}{r}(0)=0.
\end{eqnarray*}
Indeed, $\phi(r)$ is given explicitly by
\begin{eqnarray}
  \label{eq:3.7}
  \phi(r)=\theta_1+(\theta_2-\theta_1)
J_0\left(\frac{rv_0\sqrt{\alpha}}{\kappa}\right),~~~
\alpha=\frac{f(\theta_2)}{\theta_2-\theta_1}~~
\hbox{for $\displaystyle r\le R_1=\frac{\kappa\xi_1}{v_0\sqrt{\alpha}}$.}
\end{eqnarray}
Here $J_0(\xi)$ is the Bessel function of order zero, and $\xi_1$ is
its first zero. Furthermore, we have
\begin{eqnarray}
  \label{eq:3.8}
  \phi(r)=B\ln\frac{r}{R},~~ \hbox{for $R_1\le r$}
\end{eqnarray}
with $B$ and $R$ determined by matching (\ref{eq:3.7}) and
(\ref{eq:3.8}) at $r=R_1$. Then we get
\begin{eqnarray*}
  R=l\xi_1\sqrt{\frac{\theta_2-\theta_1}{f(\theta_2)}}
\exp\left[\frac{\theta_1}{(\theta_2-\theta_1)\xi_1 |J_0'(\xi_1)|}\right]=Cl,
~l=\frac{\kappa}{v_0}.
\end{eqnarray*}
Then $\phi(r)$ satisfies
\begin{eqnarray*}
  &&\kappa\Delta\phi+\frac{v_0^2}{\kappa}f(\phi)\ge 0,~~\phi(r)> 0, 0\le r< R\\
&&\phi(R)=0.\nonumber
\end{eqnarray*}
Thus we will take the critical size of plateau in the velocity profile 
to be $2R$ so that the disc of radius $R$ will fit in.

{\bf Step 2. A sub-solution.}
Let us now assume that $h\ge h_0=R$. We make a coordinate change 
\[
\xi=x-\bar ut
\]
In new coordinates we have a function 
$T(t,\xi,y)$ that solves 
\begin{eqnarray}
  \label{eq:3.12}
  T_t+A(-\bar u+u(y))T_\xi=\kappa\Delta_{\xi,y}T+
\frac{v_0^2}{\kappa}f(T)
\end{eqnarray}
with the initial data given by (\ref{eq:3.4.1}): 
\[
  T(0,\xi,y)=\left\{\begin{array}{ll}\phi(\xi,y), 
& (\xi,y)\in\Omega_0=B((0,a);h_0) \\ 0, &(\xi,y)\notin\Omega_0 \end{array}.
\right.
\]
Observe that $\phi(\xi,y)$ satisfies (\ref{eq:3.3}) inside $\Omega_0$ 
since $u(y)=\bar u$ in $\Omega_0$. Moreover, $T(t,\xi,y)\ge \phi(\xi,y)$
on $\partial\Omega_0$, where $\phi$ vanishes.  
Therefore the maximum principle implies that inside $\Omega_0$ we have
\begin{equation}\label{eq:3.12.3}
T(t,\xi,y)\ge\phi(\xi,y)~~\hbox{for all $t>0$ and $(\xi,y)\in\Omega_0$.}
\end{equation}
Note that $T_h(t,\xi,y)=T(t+h,\xi,y)$ solves (\ref{eq:3.12})
with the initial data
\begin{equation}\label{eq:3.12.2}
T_h(0,\xi,y)=T(h,\xi,y)\ge T(0,\xi,y).
\end{equation}
The inequality in (\ref{eq:3.12.2}) follows from (\ref{eq:3.12.3})
inside $\Omega_0$ and the fact that $T(t,x,y)\ge 0$ outside
$\Omega_0$.
Therefore we have 
\[
T(t+h,\xi,y)-T(t,\xi,y)\ge 0~~\hbox{for all
  $h>0$, $t>0$ and $(\xi,y)\in D$}
\]
and thus the limit
\[
\bar T(\xi,y)=\lim_{t\to\infty}T(t,\xi,y)
\]
exists since $T\le 1$. Moreover, the standard parabolic regularity implies that
$T(t,\xi,y)$ converges to $\bar T(\xi,y)$ uniformly on compact sets
together with its derivatives up to the second order. Therefore $\bar
T$ satisfies the stationary problem
\begin{equation}\label{eq:3.12.1}
(-\bar u+u(y))\bar T_\xi=\kappa\Delta_{\xi,y}\bar T+
\frac{v_0^2}{\kappa}f(\bar T).
\end{equation}
We also have 
\[
\bar T(\xi,y)> \phi(\xi,y)~~\hbox{for $(\xi,y)\in\Omega_0$}.
\]
It is easy to show using the sliding method of Berestycki and Nirenberg
\cite{BN} that for any $(\xi,y) \in D_h$
\begin{equation}\label{eq:3.13}
\bar T(\xi,y)> \phi(\xi-r,y)
\end{equation}
where the right side is any translation of $\phi$ along the $\xi$ axis.
 Indeed,  assume that
there exists the smallest (say, positive) $r$ such that $\bar T(\xi_0,y_0)=
\phi(\xi_0-r,y_0)$ at some point $(\xi_0,y_0)\in D_h$. Then
the strong maximum principle implies that $\bar T(\xi,y)= \phi(\xi-r,y)$ 
for all $(\xi,y)$ inside the translate $\Omega_r
=\Omega_0-r e_1$ of the disc $\Omega_0$. But that contradicts
the fact that $\bar T(\xi,y)> \phi(\xi-r,y)=0$ on the boundary
$\partial\Omega_r$. Then (\ref{eq:3.13}) implies that
\begin{equation}\label{eq:3.14}
\bar T(\xi,a)> \theta_2=\sup_{(\xi,y)\in\Omega_0}\phi(\xi,y), 
~\hbox{for all $\xi\in {\mathbb R}$}. 
\end{equation}
The next two lemmas show that (\ref{eq:3.12.1}) and (\ref{eq:3.14})
imply that $\bar T(\xi,y)\equiv 1$.
\begin{lemma}\label{lemma1}
Let $\bar T$ be a solution of (\ref{eq:3.12}) such that (\ref{eq:3.14}) holds.
Then we have
  \begin{equation}
    \label{eq:3.15}
    \intl_{D}d\xi dy f(\bar T(x,y))<\infty,~~~
\intl_{D}d\xi dy|\nabla\bar T|^2<\infty.
  \end{equation}
\end{lemma}
\begin{proof} In order to show that integral of $f(\bar T)$ is finite we
integrate (\ref{eq:3.12.1}) over the set $(-L+\zeta,L+\zeta)\times[0,H]$
with $L$ large and $\zeta\in[0,l]$. We get
\begin{eqnarray*}
&&A\intl_{0}^Hdy[-\bar u+u(y)][\bar T(L+\zeta,y)-\bar T(-L+\zeta,y)]=
\kappa\intl_{0}^Hdy[\bar T_\xi(L+\zeta,y)-\bar
T_\xi(-L+\zeta,y)]\\
&&+
\frac{v_0^2}{\kappa}
\intl_{0}^Hdy\intl_{-L+\zeta}^{L+\zeta}d\xi f(\bar T(\xi,y)
\end{eqnarray*}
and average this equation in $\zeta\in[0,l]$:
\begin{eqnarray*}
&&\frac Al\intl_0^ld\zeta\intl_{0}^Hdy[-\bar u+u(y)]
[\bar T(L+\zeta,y)-\bar T(-L+\zeta,y)]\\
&&=
\frac{\kappa}{l}
\intl_{0}^Hdy[\bar T(L+l,y)-\bar T(L,y)-\bar T(-L+l,y)+\bar T(-L,y)]\\
&&+\frac{v_0^2}{l\kappa}
\intl_0^ld\zeta\intl_{0}^Hdy\intl_{-L+\zeta}^{L+\zeta}d\xi f(\bar T(\xi,y).
\end{eqnarray*}
Therefore we obtain
\begin{eqnarray*}
\frac{v_0^2}{\kappa}\intl_{0}^Hdy\intl_{-L+l}^{L}d\xi f(\bar T(\xi,y))
\le \frac{4\kappa H}{l}+2HA[\bar u+\|u\|_{\infty}]
\end{eqnarray*}
for all $L$, and hence the first inequality in (\ref{eq:3.15}) holds.
In order to obtain the second inequality we multiply (\ref{eq:3.12.1})
by $T$ and perform the same integration and averaging as above. This
leads to
\begin{eqnarray}
&&\frac A{2l}\intl_0^ld\zeta\intl_{0}^Hdy[-\bar u+u(y)]
[\bar T^2(L+\zeta,y)-\bar T^2(-L+\zeta,y)]\nonumber\\
&&=
\frac{\kappa}{2l}
\intl_{0}^Hdy[\bar T^2(L+l,y)-\bar T^2(L,y)-\bar T^2(-L+l,y)+\bar T^2(-L,y)]
\nonumber\\
&&+\frac{1}{l}
\intl_0^ld\zeta\intl_{0}^Hdy\intl_{-L+\zeta}^{L+\zeta}d\xi \left[
\frac{v_0^2}{\kappa}Tf(\bar T)+\kappa|\nabla T|^2\right]\label{eq:3.12.4}
\end{eqnarray}
and then the second inequality in (\ref{eq:3.15}) follows from
(\ref{eq:3.12.4}) and the first inequality in (\ref{eq:3.15}). 
\end{proof}

\begin{lemma}\label{lemma2}
The limit function $\bar T(\xi,y)\equiv 1$
\end{lemma}
\begin{proof}
  Notice
that $\bar T$ can not achieve local minima in $D$ as follows from the
maximum principle.  Therefore if we define
$\displaystyle\mu_{\alpha,\beta}= \min_{\alpha\le \xi\le \beta, 0\le
  y\le H}\bar T(\xi,y)$ and $\displaystyle\mu(\alpha)=\min_{ 0\le y\le
  H}\bar T(\alpha,y)$ then $\mu_{\alpha,\beta}=\mu(\alpha)$ or
$\mu_{\alpha,\beta}=\mu(\beta)$. Furthermore, if $\bar T(\xi,y)=1$ at
some point, then $\bar T(\xi,y)\equiv 1$ everywhere by the strong
maximum principle. In particular, if $\mu(0)=1$, then $\bar
T(\xi,y)\equiv 1$. Therefore we consider only the case that
$\mu(0)<1$ and argue by contradiction.
We have either $\mu({\xi})<\mu(0)$ for any $\xi>0,$ or for any $\xi <0.$ 
Otherwise the minimum of $\bar T$ over the set $[-\xi,\xi]\times[0,H]$
would be achieved inside.  Let us assume without loss of generality
that $\mu(\xi)<\mu(0)$ for any $\xi>0$.  Consider 
\[ \delta = {\rm min} \left( \frac{1-\mu(0)}{2}, \frac{\theta_2-\theta_0}{2} \right). \]
For any $\xi>l,$ we have three options. \\
1.  $\bar T(\xi,y) \in [\mu(\xi), \mu(\xi) + \delta]$ for any $y \in [0,H],$
and $\mu(\xi) \geq \theta_0+\delta.$ 
In this case, by definition of $\delta,$ properties (\ref{eq:2.1.2}) 
of $f,$ and since  $\mu(\xi) \leq \mu(0)<1,$ we have
\[ \int\limits_0^H f( \bar T(\xi,y))\,dy \geq CH. \] \\
2. $\mu(\xi) <\theta_0+\delta.$ Inequality (\ref{eq:3.14}) implies that there exist $y_1,$ $y_2$ 
such that $\bar T(\xi,y_1) = \theta_2-\delta,$ $\bar T(\xi,y_2) = \theta_2$. 
Then 
\[ 
\left| \int\limits_{y_1}^{y_2} f(\bar T(\xi,y))\,dy\right| \geq C |y_2-y_1|, \]
\[  \int\limits_{y_1}^{y_2} |\bar T_y|^2 \,dy \geq |y_2-y_1|^{-1} 
\left( \int\limits_{y_1}^{y_2}\, |\bar T_y| \, dy \right)^2 \ge 
\delta^{2}|y_2-y_1|^{-1}. \]
Therefore, 
\begin{equation}
\label{eqprod}
\int\limits_{H}^{0} f(\bar T(\xi,y))\,dy  
\int\limits_{y_1}^{y_2} |\bar T_y|^2 \,dy  \geq C \delta^2 
\end{equation}
in this case. \\
3. The remaining option is that $\mu(\xi) \geq \theta_0+\delta,$ and
there exists $y \in [0,H]$ such that $\bar T(\xi,y)>\mu(\xi)+\delta.$
In this case, an argument identical to the reasoning
of option two leads to the same bound (\ref{eqprod}). \\
Overall, we see that for any $\xi<l,$
\[ 
\int\limits_0^H \left( f(\bar T(\xi,y) + |\nabla \bar T|^2(\xi,y) \right)\,
dy \geq C, 
\]
where $C$ depends only on $H$ and $\delta.$ But this contradicts
directly Lemma~\ref{lemma1}.
\end{proof}

Using Proposition~\ref{prop1}, we can now complete the proof of
Theorem~\ref{thm2}.  Notice that even though we showed $\bar T(\xi ,y)
\equiv 1,$ we still have to show the limiting function is the same in
the original coordinates.  Lemma~\ref{lemma2} implies that there
exists a time $t_0$ so that for all
$\xi\in[0,L_0(\theta_2-\theta_0,Au(y))]$ and all $y\in[0,H]$ we have
\[
T(t_0,\xi,y)\ge \theta_0+\eta.
\]
Here $L_0(\eta,v)$ is defined by Proposition \ref{prop1}. Then we may
apply Proposition \ref{prop1} with $s=0$ in the original coordinates
$(x,y)$, and initial data $T(t_0,x,y)$, and get (\ref{eq:1.2}). The
fact that $c_l<0<c_r$ follows from (\ref{eq:3.1.1}) and, for instance,
results of \cite{Ber-Lar-Lions}. 
\end{proof}

The sub-solution we constructed in  Theorem~\ref{thm2} is also useful for a 
proof of Theorem~\ref{thm3}.
As mentioned in the Introduction, we only  need to show the lower bound, the 
upper bound is contained in \cite{CKOR}. We begin the proof with two 
auxiliary lemmas.
\begin{lemma}\label{lemma3} 
Let $u(y)$ be a function that is constant on an interval:
\[
u(y)=\bar u,~~y\in[a-h,a+h]
\]
that may or may not satisfy the mean zero condition (\ref{eq:3.1.1}).
Assume that $h\ge h_0$ with $h_0$ given in Theorem \ref{thm2}. Then we
have $c_l(u) \leq \overline{u} \leq c_r(u)$ (recall that $c_l(u)$ and
$c_r(u)$ are the velocities of unique left and right traveling
fronts).
\end{lemma}
\begin{proof}
 Let $T_0$ be an initial data as in Theorem \ref{thm2} 
such that
\[
T_0(x,y)\ge 
\left\{\begin{array}{ll}\phi(x,y), 
& (x,y)\in\Omega_0=B((0,a);h_0) \\ 0, &(\xi,y)\notin\Omega_0 \end{array}
\right.
\]
with the function $\phi(x,y)$ constructed in the proof of Theorem \ref{thm2}
and given explicitly by (\ref{eq:3.7}) and (\ref{eq:3.8}). Then we have
\[
T(t,x-\bar ut,y)\ge\phi(x,y)
\]
and therefore Proposition \ref{prop1} implies that $c_l\le\bar u\le
c_r$ because $T(t,x-\bar ut,y)$ may not go to zero as $t\to\infty$. 
\end{proof}

Let $y_0$ be a point at which $|u_+|_{h_0}$
is achieved: $u(y)\ge |u_+|_{h_0}$ for  $y\in[y_0-h_0,y_0+h_0]$.
We define now a new velocity field
\begin{equation}\label{eq:3.19}
v(y)= \left\{\begin{array}{ll}u(y), & y\notin[y_0-h_0,y_0+h_0] \\ 
      |u_+|_{h_0}, & y\in [y_0-h_0,y_0+h_0]\end{array}  \right.
\end{equation}
so that $u(y)\ge v(y)$.
\begin{lemma}\label{lemma4}
  Let $u(y)\ge v(y)$ be two velocity fields. Then the corresponding
  right traveling front speeds satisfy $c_r(u)\ge c_r(v)$.
Similarly, if $u(y) \leq v(y),$ then $c_l(u) \leq c_l(v).$ 
\end{lemma}
\begin{proof}
 Let $T^{u,v}(t,x,y)$ be solutions of
\[
T_t^u+Au(y)T_x^u=\kappa\Delta T^u+\frac{v_0^2}{\kappa}f(T^u)
\]
and
\[
T_t^v+Av(y)T_x^v=\kappa\Delta T^v+\frac{v_0^2}{\kappa}f(T^v)
\]
with the same initial data
\[
T^u(0,x,y)=T^{v}(0,x,y)=T_0(x,y).
\]
The function $T_0(x,y)$ is assumed to be front-like and monotonic. That is,
$T_0(x,y)=1$ for $x\le -L$, $T_0(x,y)=0$ for $x\ge L$ and 
\[
\pdr{T_0}{x}(x,y)\le 0.
\]
Then by maximum principle applied to the equation for
${\partial T^{u,v}}/{\partial x}$ we have
\[
\pdr{T^{u,v}}{x}(t,x,y)\le 0 ~~\hbox{for all $t>0$}
\]
and hence
\[
T_t^v+Au(y)T_x^v-\kappa\Delta T^v-\frac{v_0^2}{\kappa}f(T^v)=A(u(y)-v(y))T_x^v
\le 0.
\]
Therefore by maximum principle we have
\begin{equation}\label{eq:3.20}
T^u(t,x,y)\ge T^v(t,x,y).
\end{equation}
However, by  results of \cite{jxin-2,Roq-2} similar to Proposition
\ref{prop1} but for the front-like initial data we have
\[
  \lim_{t\to\infty}T^v(x-st,y,t)=
\left\{\begin{array}{ll} 1, & \hbox{if $s<c_r(v)$} \\
0, & \hbox{if $s>c_r(v)$} \end{array}\right.
\]
and similarly for $T^u$. Then $c_r(v)>c_r(u)$ would be incompatible with
(\ref{eq:3.20}), and hence $c_r(v)\le c_r(u)$. 
Naturally, an analogous result is true for left traveling fronts. 
\end{proof}
\begin{proof}[Proof of Theorem~\ref{thm3}]
In order to finish the proof of Theorem \ref{thm3} we apply Lemma
\ref{lemma4} to $v(y)$ given by (\ref{eq:3.19}), and observe that
$c_r(v)$ satisfies the estimate in Theorem \ref{thm3} according to Lemma
\ref{lemma3}.
\end{proof}

Finally, we remark that in the situation of Theorem~\ref{thm2} strong
advection not only does not quench the flame, but has directly
opposite effect, according to Theorem~\ref{thm3} and
Proposition~\ref{prop1}.

\begin{corollary}
\label{deton}
Let $u(y)=\bar u =const$ for $y \in [a-h,a+h]$ for some $a \in [0,H],$
and $h$ is larger than or equal to the critical size $h_0.$ Then,
provided that the size of initial data $L \geq h_0,$ we have
\begin{eqnarray*}
  \lim_{t\to\infty}T(x-st,y,t)=
\left\{\begin{array}{ll} 1, & \hbox{if $c_l<s<c_r$} \\
0, & \hbox{if $c_l>s$ or $s>c_r$}\end{array}\right.
\end{eqnarray*}
Moreover, $c_r \geq |u_+|_{h_0}$ and $c_l \leq -|u_-|_{h_0}.$
\end{corollary}
\begin{proof} 
  The first statement is a direct corollary of Proposition~\ref{prop1}
  and Theorem~\ref{thm2}.  The second statement is the content of
  Theorem~\ref{thm3}.
\end{proof}

\section{A lower bound for the quenching size}

Recall that the burning rate is defined by
\[
V(t)=\intl_DT_t(x,y)\frac{dxdy}{H}.
\]
We also say that nonlinearity $f(T)$ is of concave KPP class if 
\[
f(0)=f(1)=0,~~~\hbox{$f(T)>0$ for $0<T<1$, $f''(T)<0$.}
\]
We have previously shown \cite{CKOR} that for such nonlinearities and for
front-like initial conditions the burning rate in the presence of
any advection is bounded from below by $Cv_0$, more precisely:
\begin{equation}\label{eq:6.2}
V(t)\ge Cv_0(1-e^{-Cv_0^2t/\kappa}).
\end{equation}
The physical meaning of (\ref{eq:6.2}) is that no advection may slow
up the burning significantly.  We now show that similarly there is a
fixed size of initial data that is quenched by any shear flow. That
means that no shear flow may help prevent quenching of initial data
with a fixed small support. Let $T(t,x,y)$ be solution of
\begin{equation}\label{eq:6.3}
T_t+u(y)T_x=\kappa\Delta T+\frac{v_0^2}{\kappa}f(T)
\end{equation}
with the initial data $T(0,x,y)=T_0(x,y)$ and periodic boundary
conditions (\ref{eq:2.1.1}).
\begin{prop}\label{prop:6.1}
  There exists a constant $C>0$ that depends only on the non-linearity
  $f(T)$ such that given any initial data $T_0$ with
\[
\intl_DT_0(x,y){dxdy}\le Cl^2,~~l=\frac{\kappa}{v_0},
\]
we have
\begin{equation}\label{eq:6.4}
T(t,x,y)\to 0 \hbox{~~~uniformly in $D$ as $t\to +\infty$}
\end{equation}
for all $u(y)\in C[0,H]$.
\end{prop}
\begin{proof}
We will show that solution of (\ref{eq:6.3}) with $f(T)$
replaced by a KPP nonlinearity $\tilde f(T)=MT(1-T)$ and the same
initial data drops below $\theta_0$ before time $t_0=\kappa/v_0^2$.
That will imply (\ref{eq:6.4}). Choose $M$ such that $f(T)\le \tilde
f(T)=MT(1-T)$ (such $M$ exists since $f$ is Lipschitz continuous) and
let $\tilde T$ be the solution of
\[
\tilde T_t+u(y)\tilde T_x=\kappa\Delta \tilde
T+\frac{v_0^2}{\kappa}\tilde f(\tilde T)
\]
with initial data $T_0$ and periodic boundary conditions
(\ref{eq:2.1.1}). Let us also define
\[
\tilde V(t)=\intl_D\tilde
T_t(x,y)\frac{dxdy}{H}=\frac{Mv_0^2}{\kappa}\intl_D\tilde T(1-\tilde
T)\frac{dxdy}{H}.
\]
Note that $\tilde T(t,x,y)\ge T(t,x,y)$ and hence Proposition
\ref{prop:6.1} follows from the following Lemma.
\begin{lemma}\label{lemma6.1}
  There exists a constant $C>0$ that depends only on the non-linearity
  $f(T)$ such that given any initial data $T_0$ with
\[
\intl_DT_0(x,y){dxdy}\le Cl^2,~~l=\frac{\kappa}{v_0},
\]
and any $u(x,y)\in C^1(D)$, there exists a time $t_1<t_0$ such that
$\tilde T(t,x,y)\le\theta_0$ for all $(x,y)\in D$.
\end{lemma}
\begin{proof}
The following inequality holds for $\tilde V(t)$:
\[
\tilde V(t)+\frac{\kappa}{Mv_0^2}\frac{d\tilde V(t)}{dt}\ge
2\kappa\intl_D|\nabla \tilde T|^2\frac{dxdy}{H}.
\]
We define the set
\[
{\mathcal S}(t)=\left\{y\in[0,H]:~\exists
x\in{\mathbb R}~\hbox{such that}~~T(t,x,y)\ge \theta_0/e^{2M}\right\}\subset
[0,H].
\]
Observe that there exists a constant $\delta>0$ that depends only on
$M$ and $\theta_0$ such that if the Lebesgue measure $|{\mathcal
  S}(\tau)|\le \delta l$ at some time $\tau\in[0,t_0/2]$, then
$\|\tilde T(t_0)\|_{L^\infty}\le\theta_0$. Indeed, we have then
\[
\tilde T(\tau,x,y)\le \chi(y)+\psi(x,y),~~~
\|\chi(y)\|\le \delta l,~~~\psi(x,y)\le\theta_0/e^{2M}.
\]
Therefore we have with $C$ as in (\ref{eq:simpleest})
\[
T(t_0,x,y)\le e^M\left[(C+1)\delta+\frac{\theta_0}{e^{2M}}\right]\le \theta_0
\]
for a sufficiently small $\delta$. Hence it suffices to consider the
case when $|{\mathcal S}(t)|\ge \delta l$ for all $t\in[0,t_0/2]$.  We
claim that then at any time $t\in[0, t_0/2]$ we have
\begin{equation}\label{eq:6.7}
\intl_D|\nabla \tilde T|^2{dxdy}\intl_D\tilde f(\tilde T){dxdy}\ge Cl^2
\end{equation}
with the constant $C$ depending only on $M$ and $\theta_0$. This may
be proved similarly to Lemma 2 in \cite{CKOR}.  Therefore we obtain
\[
\tilde V(t)+\frac{\kappa}{Mv_0^2}\frac{d\tilde V(t)}{dt}\ge
\frac{2Cv_0^2l^2}{H^2\tilde V(t)}.
\]
Thus we have
\[
\tilde V(t)\ge \frac{Cv_0l}{H}~~\hbox{for $t_0/2\le t\le t_0$}.
\]
Therefore 
\[
\intl_D\tilde T(t_0,x,y)\ge  \intl_D T_0(x,y)dxdy+C\kappa t_0.
\]
However, we have an a priori bound
\[
\intl_D\tilde T(t_0,x,y)dxdy\le e^{M}\intl_D T_0(x,y)dxdy
\]
which contradicts the previous inequality if
\begin{equation}\label{eq:6.8}
\intl_D T_0(x,y)dxdy\le\frac{C\kappa t_0}{e^M-1}=C_Ml^2.
\end{equation}
Therefore $T(t,x,y)$ has to drop below $\theta_0$ if the initial data
satisfies (\ref{eq:6.8}) and Lemma \ref{lemma6.1} is proved.
\end{proof}
Proposition \ref{prop:6.1} then follows from Lemma \ref{lemma6.1}.
\end{proof}

{\it Remark.} The uniform quenching size in Proposition \ref{prop:6.1}
is optimal. Indeed, the subsolution we have constructed in Section
\ref{sec:propagate} for shear flows with a flat part has $L^1$ norm
$Cl^2$, and thus one cannot expect that initial data with $L^1$-norm
larger than $Cl^2$ are quenched by all shear flows.

\end{document}